\documentclass[preprint]{aastex}
\begin{document}
\title{The robustness of the galaxy distribution function to effects of merging
and evolution}
\author{Abel Yang}
\affil{Department of Astronomy, University of Virginia, Charlottesville, VA
22904}
\author{William C. Saslaw}
\affil{Institute of Astronomy, Madingley Road, Cambridge CB3 0HA, UK; and
Department of Astronomy, University of Virginia, Charlottesville, VA 22904}
\author{Aik Hui Chan}
\affil{Department of Physics, National University of Singapore, 2 Science
Drive 3, Singapore 117542; and Institute of Advances Studies, Nanyang
Technological University, \#02-18 60 Nanyang View, Singapore 639673}
\author{Bernard Leong}
\affil{Senatus Pte. Ltd, 19 Amoy Street, 02-01, Singapore 069584}
\begin{abstract}
We examine the evolution of the spatial counts-in-cells distribution of galaxies
and show that the form of the galaxy distribution function does not change
significantly as galaxies merge and evolve. In particular, bound merging pairs
follow a similar distribution to that of individual galaxies. From the adiabatic
expansion of the universe we show how clustering, expansion and galaxy mergers
affect the clustering parameter $b$. We also predict the evolution of $b$ with
respect to redshift.
\end{abstract}

\keywords{galaxies: statistics --- cosmology: theory --- large-scale structure
of universe}
%\bodymatter
\section{Introduction}
\label{sec-intro}
The galaxy spatial distribution function $f(N,V)$ is a simple but powerful
statistic which characterizes the locations of galaxies in space. It includes
statistical information on voids and other underdense regions, on clusters of
all shapes and sizes, on filaments, on the probability of finding an arbitrary
number of neighbors around randomly located positions, on counts of galaxies in
cells of arbitrary shapes and sizes randomly located, and on galaxy correlation
functions of all orders. These are just some of its representations
~\citep{2000gpsg.book.....S, 2009arXiv0902.0747S}. Moreover it is also closely
related to the distribution function of the peculiar velocities of galaxies
around the Hubble flow~\citep{1990ApJ...365..419S, 2004ApJ...608..636L}.

A physically motivated form of the distribution function for galaxies in
quasi-equilibrium was derived and generalized using a statistical mechanical
approach.~\citep{2002ApJ...571..576A, 2006ApJ...645..940A} It has also been
generalized to systems containing particles of two different
masses~\citep{2006IJMPD..15.1267A}. In its simplest form, the probability of
finding $N$ galaxies in a cell of volume $V$ is given by the gravitational
quasi-equilibrium distribution(GQED):
\begin{equation}\label{eint-GQED}
f_V(N) = \frac{\overline{N}(1-b)}{N!} \left(\overline{N}(1-b)+Nb\right)^{N-1}
\exp\left(-\overline{N}(1-b)-Nb\right)
\end{equation}
where $\overline{N}$ is the average number of galaxies in a cell and the
clustering parameter $b=-W/2K$ is the ratio of the gravitational correlation
energy $W$ to twice the kinetic energy $K$ of peculiar velocities relative to
the Hubble flow. Although other distributions have been proposed~(e.g. negative
binomial; for an early review see \citealt{1986ApJ...306..358F}), they are
generally not physically motivated. Extensive computer simulations~(summarized
in \citealt{2000gpsg.book.....S}) designed to test the GQED agree closely
with the analytical results. Observations at both low~\citep{2005ApJ...626..795S}
and high~\citep{2009ApJ...695.1121R} redshifts suggest that the form of the
distribution function $f_V(N)$ is essentially unchanged over a wide range of
redshifts. However, mergers among galaxies could have modified the form of
$f_V(N)$ significantly compared to the simple model which involves no mergers at
all, and yet they did not. In this paper we examine the robustness of the GQED
to merging galaxies.

\section{The positions of merging galaxy pairs}
\label{sec-merge}
Merging galaxy pairs are often extended structures that would have a different
interaction potential than simple spheres or point masses. With a different
interaction potential, they may also be distributed differently. Extended
structures resulting from mergers will not only have a different interaction
potential, they also have a different mass and should be treated differently. To
consider how these extended structures influence the GQED, we first consider how
they modify the interaction potential. Using the modified potential we can then
derive the distribution function for an ensemble of species with a range of
masses.

The generalized interaction potential between particles (galaxies) each of
which has an isothermal halo is given by~\citep{2002ApJ...571..576A}
\begin{equation} \label{emp-isohalo}
\phi(r) = -\frac{G m^2}{(r^2+\epsilon^2)^{1/2}}
\end{equation}
where $G$ is the gravitational constant, $r$ is the separation between a pair of
particles, $m$ is the mass of each particle, and $\epsilon$ is a parameter
related to the finite size of a galaxy in proper coordinates. We can generalize
this interaction potential by factoring out the $\epsilon$ terms to get
\begin{equation} \label{emp-halo}
\phi(r) = -\frac{G m^2}{r} \kappa(\epsilon/r)
\end{equation}
where $\kappa(\epsilon/r)$ represents a modification to the Newtonian potential.
This modification only affects the potential energy part of the configuration
integral given by equation (3) of \citet{2002ApJ...571..576A}
\begin{equation} \label{emp-config}
 Q_N(T,V) =
\int\ldots\int\exp\left[-\phi(\mathbf{r}_1,\mathbf{r}_2,\ldots,
\mathbf{r}_n)T^{-1}\right] d^{3N}\mathbf{r}
\end{equation}
where $T$ is the kinetic temperature of the ensemble in units where Boltzmann's
constant is unity, and $\phi$ is the interparticle potential energy of the
ensemble. By following the procedure in section 2 of
\citet{2002ApJ...571..576A}, the potential energy part of the Hamiltonian
for a 2-galaxy system becomes
\begin{equation} \label{emp-q2e}
Q_2(T,V) = V^2\left[1+\frac{3Gm^2}{2T(\overline{n})^{-1/3}}
\zeta\left(\frac{\epsilon}{R_1}\right)\right ]
\end{equation}
where $R_1$ is the scale where the two-galaxy correlation function is
negligible, $\overline{n}$ is the number of particles per unit volume, and
\begin{equation} \label{emp-zeta}
\zeta\left(\frac{\epsilon}{R_1}\right) =
\int_0^{R_1}\frac{2r}{R_1^2}\kappa\left(\frac{\epsilon}{r}\right) dr
\end{equation}
describes how a modification to the potential changes the partition function.

The modification given by equation (\ref{emp-zeta}) is analogous to
$\alpha(\epsilon/R_1)$ given by equation (16) of
\citet{2002ApJ...571..576A}, but can describe a generalized modification
of the potential rather than the particular modification that arises from an
isothermal halo. The effects of a modified potential enter into the distribution
function only through the parameter $b$, which now becomes
\begin{equation}
\label{emp-bdef}
b_\epsilon = \frac{(3/2)G^3m^6 \overline{n} T^{-3}
\zeta(\epsilon/R_1)}{1+(3/2)G^3m^6 \overline{n} T^{-3} \zeta(\epsilon/R_1)}.
\end{equation}
For an attractive potential, $\kappa(\epsilon/r)$ is is always positive and
hence $\zeta(\epsilon/R_1) > 0$ for all values of $R_1$ so from equation
(\ref{emp-bdef}), $0\leq b_\epsilon \leq 1$.

From the thermodynamic variables of the system given by equations (26)-(30) of
\citet{2002ApJ...571..576A}, we see that the forms of the distribution
function $f_V(N)$ and the thermodynamic functions of the system are essentially
unchanged by a modified potential although their values are different. This
shows that the form of the galaxy distribution function is robust to modified
potentials. This also allows us to treat merging galaxy pairs as a separate
species and extend the GQED to describe the distribution of merging galaxy
pairs.

A merging galaxy pair which is bound will have its dynamical center at its
center-of-mass, and will relax to form a single galaxy at its center-of-mass.
Hence the position of the centers-of-mass of these merging pairs will be related
to the positions of the merged galaxies. While they merge, these merging pairs
form an extended system with an external potential given by the vector sum of
the external potentials of both galaxies in the merging pair:
\begin{equation} \label{emg-ppot}
\phi_{2}(\mathbf{r}) = -\frac{Gm_1M}{\|\mathbf{r}+\mathbf{x}_1\|}
-\frac{Gm_2M}{\|\mathbf{r}+\mathbf{x}_2\|}
= -\frac{Gm^2}{\|\mathbf{r}\|}
\left(\frac{\|\mathbf{r}\|}{\|\mathbf{r}+\mathbf{x}_1\|}
+\frac{\|\mathbf{r}\|}{\|\mathbf{r}+\mathbf{x}_2\|}\right)
\end{equation}
where $\mathbf{r}$ is the distance from the center of mass of the merging pair
to a more distant galaxy of mass $M$, and $\mathbf{x}_1$ and $\mathbf{x}_2$ are
the distances from each component to the center-of-mass of the system. To
approximate the case when mergers between similar-sized galaxies have the
dominant effect on the distribution function, we assume that all galaxies have
the same mass $m$ so that $m = m_1 = m_2 = M$. From equation (\ref{emg-ppot}),
the first order approximation to the external potential contains a factor of
$2m$. In the case where $\|\mathbf{r}\| \gg \|\mathbf{x}\|$, the first order
term dominates and we obtain the modification factor $\kappa_2 \approx 2$ to the
potential that arises from treating these merging pairs as single extended
particles.

The universe however does not contain only merging pairs. To model the presence
of individual galaxies that are not currently merging, we consider a two-species
distribution~\citep{2006IJMPD..15.1267A} where one species consists of merging
pairs, and the other species consists of galaxies that are not merging. For
simplicity we consider a system containing two species of different extended
particles where species 1 represents individual galaxies with an average mass of
$m$, each with a halo, and species 2 represents bound pairs of merging galaxies
with a total mass of $2m$. Using the modification to the GQED described above,
the potential between a pair of particles of species 1 (each of which is a
galaxy) is
\begin{equation} \label{emg-phihalo}
\phi_1(r) = \frac{G m_1^2}{r} \kappa_1(\epsilon_1/r)
\end{equation}
where $\kappa_1(\epsilon_1/r)$ is a softening factor that arises from an
extended halo with a physical extent described by $\epsilon_1$~(e.g. isothermal
halo~\citep{2002ApJ...571..576A}). Likewise, the external potential between a
particle of species 1 (a galaxy) and a particle of species 2 (a bound merging
pair) is given by \begin{equation} \label{emg-phipair}
\phi_2(r) = \frac{G m_1 m_2}{r} \kappa_2(\epsilon_2/r)
\end{equation}
where $m_2 = 2 m_1$ because there are two galaxies of mass $m_1$ in each merging
pair, and the modification to the potential is given by
$\kappa_2(\epsilon_2/r)$. Since a modified potential only changes $b$ in the
single-species distribution function, a modified potential only changes the
two-species clustering parameter $b_m$ in the two-species distribution function
\begin{equation} \label{emg-bm}
b_{m}
 = \frac{N_{1}}{N}
\frac{\beta\overline{n}T^{-3}}{1+\beta\overline{n}T^{-3}}+\frac{N_{2}}{N}
\frac{\beta_{12}\overline{n} T^{-3}}{1+\beta_{12}\overline{n} T^{-3}}
 = \frac{b}{1+N_{2}/N_{1}}\left(1+\frac{(N_{2}/N_{1})(\beta_{12}/\beta)}
{1-b+(\beta_{12}/\beta)b}\right)
\end{equation}
where $\beta$ and $\beta_{12}$ are given by
\begin{eqnarray} \label{emg-beta}
\beta &=& \frac{3}{2}(Gm_1^2)^3\zeta_{2}\left(\frac{\epsilon_1}{R_1}\right) =
\frac{3}{2}(Gm_1^2)^3\int_0^{R_1}\frac{2r}{R_1^2}\kappa_{1}\left(\frac{
\epsilon_1}{r}\right)dr \\
\beta_{12}&=&\frac{3}{2}(Gm_1m_2)^3\zeta_{2}\left(\frac{\epsilon_2}{R_1}\right) =
\frac{3}{2}(Gm_1m_2)^3\int_0^{R_1}\frac{2r}{R_1^2}\kappa_{2}\left(\frac{
\epsilon_2}{r}\right)dr
\end{eqnarray}
and $b$ is the single-species clustering parameter given
by~\citep{2002ApJ...571..576A}
\begin{equation} \label{emp-b}
b = \frac{\beta \overline{n} T^{-3}}{1+\beta \overline{n} T^{-3}}.
\end{equation}

The two-species distribution function is thus~\citep{2006IJMPD..15.1267A}
\begin{eqnarray} \label{emg-df}
f_{V}(N) &=& \frac{\overline{N}(1-b)}{N!} \left[\overline{N}(1-b)+N
b\right]^{N_{1}-1} \left[\frac{\overline{N}(1-b)+(\beta_{12}/\beta) Nb}
{1-b+(\beta_{12}/\beta) b}\right]^{N_{2}} \nonumber \\
 & & \times \exp[-\overline{N}(1-b_{m})-Nb_{m}].
\end{eqnarray}
The two-species distribution reduces to the single-species distribution function
in the limit $N_{2}/N_{1} \to 0$ and $\beta_{12}/\beta \to 1$. When $N_1 \gg
N_2$ and $\overline{N}$ is large, the deviation from the GQED is small because
the $\left[\overline{N}(1-b)+Nb\right]^{N_1-1}$ term dominates. Measurements of
the fraction of merging pairs in the VVDS catalog~\citep{2009A&A...498..379D}
suggest that for redshifts of $z \lesssim 1$, $N_2/N_1\approx 10\%$. Assuming
these mergers are between galaxies that are similar in mass, $m_2 \approx 2 m_1$
so $\beta_{12}/\beta \approx 8$. For large cells which are a representative
sample of the universe, $\overline{N} \gtrsim 100$ and the difference between
the two-species distribution and the single-species distribution is small on the
level of about $5\%$. Hence under these conditions, the only significant effect
of galaxy mergers in this context is a change in the average mass of a galaxy.

\section{Redshift evolution of $b$}
\label{sec-dbdz}
To determine the change in the clustering parameter $b$ we consider a merging
pair of galaxies each of mass $m$ which approach each other with velocities of
$\mathbf{v}_{1}$ and $\mathbf{v}_{2}$. Since momentum is conserved, the merged
galaxy follows the trajectory of the center of mass of the progenitors, and has
a final velocity after the merger of $\mathbf{v}_{f} = (\mathbf{v}_{1} +
\mathbf{v}_{2})/2$. The final velocity of the merged galaxy depends on the
detailed dynamics of the system, but by averaging over all orientations, we find
that mergers will not change the average kinetic energy of an ensemble. Hence
the more important contribution to the evolution of $b$ comes from the change in
the positions of galaxies and the expansion of the universe.

We extend the analysis of \citet{1992ApJ...391..423S} to describe an
ensemble where galaxies merge by considering the effect of the adiabatic
expansion of the universe. The equations of state for internal energy $U$ and
pressure $P$ are~\citep{2000gpsg.book.....S, 2002ApJ...571..576A}
\begin{equation} \label{edb-U}
U = \frac{3}{2}\overline{N}T(1-2b)
\end{equation}
and
\begin{equation} \label{edb-P}
P = \frac{\overline{N}T}{V}(1-b)
\end{equation}
where $V$ is the volume and $T$ is the kinetic temperature of peculiar
velocities in units where the Boltzmann constant is $1$. Equations
(\ref{emp-b}) and (\ref{edb-P}) can be combined to get~\citep{1992ApJ...391..423S}
\begin{equation} \label{edb-b0}
b=\frac{b_0 \overline{n} T^{-3}}{1+ b_0 \overline{n} T^{-3}}=b_0 P T^{-4}.
\end{equation}
The analysis of section \ref{sec-merge} generalizes equation (\ref{edb-b0}) to
extended objects with the generalized form of $b$ given in equation
(\ref{emp-bdef}). By comparing equations (\ref{edb-b0}) and (\ref{emp-bdef}) we
see that $b_0$ is given by
\begin{equation} \label{edb-b0def}
b_0 = \frac{3}{2} G^3\overline{m}^6 \zeta\left(\frac{\epsilon}{R_1}\right)
\end{equation}
where $\overline{m}$ is the average mass of a galaxy, and $\zeta$ is a function
of order unity that depends weakly on $\epsilon/R_1$.

Because $R_1$ is defined as the scale where the two-galaxy correlation function
becomes negligible, the universe is approximately uniform averaged over scales
$\gtrsim R_1$. Here we take $R_1$ to be the scale at which the two-galaxy
correlation function begins to decrease faster than a power law. Measurements
from the 2DFGRS~\citep{2003MNRAS.346...78H} have indicated that $R_{1}$ is about
$12h^{-1}$ Mpc at which the two-galaxy correlation function is of the order
$10^{-2}$. Here $h = H_0/100$ is the reduced Hubble constant. We note that such
cells are large enough to contain individual field galaxies and clusters of
galaxies, and hence would be an approximately representative sample of the
universe. Assuming that on such scales, galaxies have isotropic average
velocities, then for cells with a radius larger than $R_{1}$, galaxies are as
likely to enter a cell as they are to leave a cell. With this assumption, the
total mass in each comoving cell $M_{c}$ would be approximately constant, and
each cell would have on average $\overline{N} = M_{c}/\overline{m}$ galaxies and
$d(\overline{m}\overline{N})=0$. Therefore $\overline{N} \propto
\overline{m}^{-1}$ and $b_0$ can be written as a function of $\overline{N}$
instead of $\overline{m}$. This transforms equation (\ref{edb-b0}) into the form
\begin{equation} \label{edb-b0P}
b = b_0(\overline{N}) P T^{-4} = \frac{3}{2}
G^3\left(\frac{M_c}{\overline{N}}\right)^6
\zeta\left(\frac{\epsilon}{R_1}\right) P T^{-4}.
\end{equation}
Differentiating equation (\ref{edb-b0P}) with respect to $\overline{N}$ gives
\begin{equation} \label{edb-dbgen}
\frac{db}{b} = -6\frac{d\overline{N}}{\overline{N}} \left(
1+\frac{\overline{N}}{6R_1} \left(\frac{\partial\ln\zeta(\epsilon/R_1)}
{\partial\overline{N}}\right)_{T,P} \right) =
-6\frac{d\overline{N}}{\overline{N}}(1+\zeta_\star)
\end{equation}
from which we define the term
\begin{equation} \label{edb-zetastargen}
\zeta_\star = \frac{\overline{N}}{6 R_1}
\left(\frac{\partial\ln\zeta(\epsilon/R_1)} {\partial\overline{N}}\right)_{T,P}.
\end{equation}
In general, we do not rule out the possibility that $\zeta(\epsilon/R_1)$ may
indirectly depend on $\overline{N}$, hence the
$\partial\ln\zeta(\epsilon/R_1)/\partial\overline{N}$ factor in $\zeta_\star$
may be nonzero.

In the case of adiabatic expansion, equations (\ref{edb-U}) and (\ref{edb-P}),
give
\begin{equation} \label{edb-ade}
0 = dU + PdV = \frac{3}{2}(1-2b) \left[ T d\overline{N} +
\overline{N}dT|_{\overline{N},P} \right]
-3\overline{N}Tdb+\overline{N}T(1-b)\frac{dV}{V}.
\end{equation}
Equation (\ref{edb-b0P}) implies
\begin{equation} \label{edb-dt}
dT|_{\overline{N},P} = -\frac{T db}{4b} \newline
\end{equation}
and hence using $dV/V = 3dR/R$ where $R$ is the scale length of the universe, we
have
\begin{equation} \label{edb-ade2}
0 = \frac{3}{2}(1-2b)T d\overline{N}-\frac{3}{2}(1-2b)\overline{N}T
\frac{db}{4b}-3\overline{N} Tdb+3\overline{N}T(1-b)\frac{dR}{R}.
\end{equation}
Rearranging the terms and using equation (\ref{edb-dbgen}), we find in terms of
redshift $z \propto 1/R - 1$
\begin{equation} \label{edb-dbdz}
\frac{db}{dz} = -\frac{1-b}{1+z}\left(\frac{1+6b}{8b}
 +\frac{1-2b}{12b \zeta_\star}\right)^{-1}.
\end{equation}

To illustrate how mergers contribute to the time evolution of $b$, we consider a
simple model of a galaxy. In our model, galaxies have isothermal halos with a
characteristic radius $\epsilon$, and all galaxies have the same density so that
in a cell of total mass $M_c$
\begin{equation} \label{edb-epsdef}
\epsilon = a\left(\overline{m}\right)^{1/3} =
a\left(\frac{M_{c}}{\overline{N}}\right)^{1/3}
\end{equation}
for some constant of proportionality $a$ such that $\epsilon$ depends on the
average mass of a galaxy. We use the GQED and form of $\zeta$ for such a case
from \citet{2002ApJ...571..576A} to obtain the constraint $-1/18 \leq
\zeta_\star \leq 0$. The extremes of this constraint gives 
\begin{equation} \label{edb-dbdz0}
\frac{db}{dz} = -\frac{1-b}{1+z} \left(\frac{24b}{5+14b} \right)
\end{equation}
for the case where $\zeta_\star = 0$ and 
\begin{equation} \label{edb-dbdzinf}
\frac{db}{dz} = -\frac{1-b}{1+z} \left(\frac{136b}{29+78b} \right)
\end{equation}
for the case with $\zeta_\star = -1/18$\ with all other cases occurring in
between equations (\ref{edb-dbdz0}) and (\ref{edb-dbdzinf}).

Since $0 \leq b \leq 1$, we compare the two cases numerically and find that the
difference between the two cases is small at less than $2\%$. This result tells
us that although galaxy mergers can influence the time evolution of $b$, their
influence mostly results from changes of the number of galaxies in a cell.

\section{Conclusion and Future Work}
\label{sec-con}

We have established that the effects of galaxy mergers leave the form of the
galaxy distribution function essentially unchanged and just alter the parameters
of the counts in cells distribution. In particular, by describing bound merging
pairs as objects with a modified gravitational potential, we obtain a modified
form of the two-species counts in cells distribution~\citep{2006IJMPD..15.1267A}
and show that it only changes the counts in cells distribution slightly from
the single-species result given by equation (\ref{eint-GQED}).

As a result of mergers, the clustering parameter $b$ increases with time, and we
have shown that it depends very weakly on the physical extent of a galaxy and
the scale $R_1$ at which the two point correlation function is negligible. The
effect of the physical extent of a galaxy changes $db/dz$ by less than 2\%,
which shows that the evolution of $b$ depends mainly on the adiabatic expansion
of the universe and the change in the number of galaxies from mergers.

These results show that even when we take galaxy mergers into account, we can
not only reproduce the GQED but also trace the evolution of the clustering
parameters. However, an analysis of the GOODS catalog~\citep{2009ApJ...695.1121R}
indicates a large variation between the North and South fields and suggests that
the sample is probably too small to draw any meaningful conclusions about the
evolution of $b$ at high redshift. Future surveys however may provide
sufficiently large samples at high redshifts to test our predicted evolution of
$b$.

\section*{Acknowledgments}

A. Yang is grateful for the support from the National University of Singapore
and the Institute of Astronomy of the University of Cambridge where part of this
work was done.

A. H. Chan would like to thank the department of History \& Philosophy of
Science, Cambridge University and Nanyang Polytechnic for kind hospitality where
part of the initial work was done.
\bibliography{paper}

\begin{thebibliography}{13}
\expandafter\ifx\csname natexlab\endcsname\relax\def\natexlab#1{#1}\fi

\bibitem[{{Ahmad} {et~al.}(2006{\natexlab{a}}){Ahmad}, {Malik}, \&
  {Masood}}]{2006IJMPD..15.1267A}
{Ahmad}, F., {Malik}, M.~A., and {Masood}, S. 2006{\natexlab{a}}, Intl. J.
  Modern Physics D, 15, 1267

\bibitem[{{Ahmad} {et~al.}(2002){Ahmad}, {Saslaw}, \&
  {Bhat}}]{2002ApJ...571..576A}
{Ahmad}, F., {Saslaw}, W.~C., and {Bhat}, N.~I. 2002, Astrophys. J., 571, 576

\bibitem[{{Ahmad} {et~al.}(2006{\natexlab{b}}){Ahmad}, {Saslaw}, \&
  {Malik}}]{2006ApJ...645..940A}
{Ahmad}, F., {Saslaw}, W.~C., and {Malik}, M.~A. 2006{\natexlab{b}}, Astrophys.
  J., 645, 940

\bibitem[{{de Ravel} {et~al.}(2009){de Ravel}, {Le F{\`e}vre}, {Tresse},
  {Bottini}, {Garilli}, {Le Brun}, {Maccagni}, {Scaramella}, {Scodeggio},
  {Vettolani}, {Zanichelli}, {Adami}, {Arnouts}, {Bardelli}, {Bolzonella},
  {Cappi}, {Charlot}, {Ciliegi}, {Contini}, {Foucaud}, {Franzetti},
  {Gavignaud}, {Guzzo}, {Ilbert}, {Iovino}, {Lamareille}, {McCracken},
  {Marano}, {Marinoni}, {Mazure}, {Meneux}, {Merighi}, {Paltani}, {Pell{\`o}},
  {Pollo}, {Pozzetti}, {Radovich}, {Vergani}, {Zamorani}, {Zucca}, {Bondi},
  {Bongiorno}, {Brinchmann}, {Cucciati}, {de La Torre}, {Gregorini}, {Memeo},
  {Perez-Montero}, {Mellier}, {Merluzzi}, \& {Temporin}}]{2009A&A...498..379D}
{de Ravel}, L. et~al. 2009, Astronomy \& Astrophysics, 498, 379

\bibitem[{{Fry}(1986)}]{1986ApJ...306..358F}
{Fry}, J.~N. 1986, Astrophys. J., 306, 358

\bibitem[{{Hawkins} {et~al.}(2003){Hawkins}, {Maddox}, {Cole}, {Lahav},
  {Madgwick}, {Norberg}, {Peacock}, {Baldry}, {Baugh}, {Bland-Hawthorn},
  {Bridges}, {Cannon}, {Colless}, {Collins}, {Couch}, {Dalton}, {De Propris},
  {Driver}, {Efstathiou}, {Ellis}, {Frenk}, {Glazebrook}, {Jackson}, {Jones},
  {Lewis}, {Lumsden}, {Percival}, {Peterson}, {Sutherland}, \&
  {Taylor}}]{2003MNRAS.346...78H}
{Hawkins}, E. et~al. 2003, Monthly Notices of the Royal Astronomical Society,
  346, 78

\bibitem[{{Leong} \& {Saslaw}(2004)}]{2004ApJ...608..636L}
{Leong}, B. and {Saslaw}, W.~C. 2004, Astrophys. J., 608, 636

\bibitem[{{Rahmani} {et~al.}(2009){Rahmani}, {Saslaw}, \&
  {Tavasoli}}]{2009ApJ...695.1121R}
{Rahmani}, H., {Saslaw}, W.~C., and {Tavasoli}, S. 2009, Astrophys. J., 695,
  1121

\bibitem[{{Saslaw}(1992)}]{1992ApJ...391..423S}
{Saslaw}, W.~C. 1992, Astrophys. J., 391, 423

\bibitem[{{Saslaw}(2000)}]{2000gpsg.book.....S}
---. 2000, {The distribution of the galaxies : gravitational clustering in
  cosmology} (Cambridge, UK: Cambridge University Press)

\bibitem[{{Saslaw} {et~al.}(1990){Saslaw}, {Chitre}, {Itoh}, \&
  {Inagaki}}]{1990ApJ...365..419S}
{Saslaw}, W.~C., {Chitre}, S.~M., {Itoh}, M., and {Inagaki}, S. 1990,
  Astrophys. J., 365, 419

\bibitem[{{Saslaw} \& {Yang}(2010)}]{2009arXiv0902.0747S}
{Saslaw}, W.~C. and {Yang}, A. 2010, in {Lecture Notes of the Les Houches
  Summer School: Long-Range Interacting Systems}, ed. T.~{Dauxois}, S.~{Ruffo},
  \& L.~F. {Cugliandolo}, Vol.~XC (Oxford, UK: Oxford University Press),
  377--398

\bibitem[{{Sivakoff} \& {Saslaw}(2005)}]{2005ApJ...626..795S}
{Sivakoff}, G.~R. and {Saslaw}, W.~C. 2005, Astrophys. J., 626, 795

\end{thebibliography}

\end{document}